\documentclass[prl,superscriptaddress,twocolumn,showpacs,preprintnumbers,amsmath,amssymb]{revtex4}

\usepackage{graphicx}
\usepackage{dcolumn}
\usepackage{bm}

\newcommand{\nl}{\newline}
\newcommand{\be}{\begin{equation}}
\newcommand{\ee}{\end{equation}}
\newcommand{\up}{\uparrow}
\newcommand{\down}{\downarrow}
\newcommand{\lag}{\langle}
\newcommand{\rag}{\rangle}

\begin{document}


\title{Phase Fluctuations in Strongly Coupled $d$-wave Superconductors}

\author{Matthias Mayr}
\affiliation{Max-Planck-Institut f\"ur Festk\"orperforschung, 70569 Stuttgart, Germany.}
\author{Gonzalo Alvarez}
\affiliation{Computational Sciences and Mathematics Division, Oak Ridge National Lab, TN 37831, USA}
\author{Cengiz \c{S}en}
\affiliation{Department of Physics, Florida State University, Tallahassee, FL 32310, USA}
\author{Elbio Dagotto}
\affiliation{Department of Physics and Astronomy, The University of Tennessee, Knoxville,
Tennessee 37996 and Condensed Matter Sciences Division, Oak Ridge National Lab, TN 37381, USA}


\begin{abstract}
We present a numerically exact solution for the BCS Hamiltonian at any temperature, 
including the degrees of freedom
associated with classical phase, as well as amplitude, 
fluctuations via a Monte Carlo (MC) integration.
This allows for an investigation over the whole range of couplings: from weak attraction,
as in the well-known BCS limit, to the mainly unexplored strong-coupling
regime of pronounced phase fluctuations. In the latter, for the first time two characteristic temperatures
$T^\star$ and $T_c$, associated
with short- and long-range ordering, respectively, can easily be identified in a mean-field-motivated Hamiltonian. 
$T^\star$ at the same time corresponds to the opening of a gap in the excitation spectrum.
Besides introducing a novel procedure to study strongly coupled $d$-wave superconductors, our results indicate that classical phase fluctuations are not sufficient to 
explain the pseudo-gap features of high-temperature superconductors (HTS).
 
\end{abstract}
\maketitle

One of the most fascinating aspects of the HTS is that a theoretical description in traditional
BCS terms -- using Cooper pairs -- is feasible, yet in many other
aspects these materials seem to deviate considerably from the standard BCS behavior.
Most notorious in this respect is the curious ``pseudogap'' (PG) phase in the underdoped regime.
The PG has attracted enormous interest in recent years and its effects have been studied 
using a wide variety of techniques
\cite{Corson_1},\cite{Lit_PG}.
It is identified as a dip in the density of states $N$$(\omega)$  
below a temperature $T^\star$, which is higher than the superconducting (SC) critical temperature $T_c$,
and its presence is
sometimes attributed to a strong coupling between the charge carriers and accompanying phase fluctuations \cite{Emery_1}. If this is the case,
then conventional mean-field (MF) methods should  not work in describing
the cuprates, since they cannot distinguish between $T^\star$ and $T_c$. For this reason,
more elaborate techniques such as diagrammatic resummations or Quantum MC
approximations have been used to address the many puzzling questions of strongly coupled superconductors.
While for the case of superconductivity with 
$s$-wave symmetry (sSC) this effort can be carried out with
the attractive Hubbard model \cite{Keller_1},
the direct study of phase fluctuations 
for $d$-wave superconductors (dSC) remains a challenge. To our knowledge, in the vast
literature on cuprates there is no available model where the physics of a strongly coupled dSC 
with short coherence lengths and large phase fluctuations 
can be studied accurately, with nearly exact solutions \cite{Nazarenko}. 
From the theory perspective, this is a conspicuous bottleneck in the HTS arena.
 
Here, we introduce a novel and simple approach to alleviate this problem. The proposed method 
allows for an unbiased treatment
of phenomena associated with classical (thermal) phase fluctuations and non-coherent pair-binding.
It represents an extension of the original solution of the pairing Hamiltonian and
has been made possible mostly due to the advance of
computational resources in the past decade.
The focus is on the
more interesting and important case - at least as far as HTS are concerned -
of a nearest-neighbor (n.n.) attraction, necessary for dSC.
With regards to the cuprates, this approach is only meaningful to the extent that the relevant phase fluctuations are 
thermal rather than quantum mechanical
and in fact it has been argued \cite{Emery_1} that phase fluctuations in cuprates may be assumed as
predominantly classical, with quantum (dynamical) fluctuations \cite{Kwon_1} suppressed. 


Our approach is built on the insight that Hamiltonians that are quadratic
in fermionic operators can be efficiently studied
with the help of Monte Carlo techniques, as has been demonstrated
in particular for the ``double-exchange'' model \cite{Dagotto_1}.
This is possible here because the original interacting model has been stripped of
quantum fluctuations in the pairing approximation.
The Hamiltonian $H_{\rm SC}$ describes an effective attraction between
fermions on a 2D lattice and 
is given by  
\begin{eqnarray}
H_{\rm SC} & = & -t\sum_{{\bf i},\delta,\sigma} c^\dagger_{{\bf i}\sigma}c^{}_{{\bf i}+\delta\sigma}+V\sum_{{\bf i},\delta}|\Delta^\delta_{\bf i}|^2-\mu\sum_{\bf i} n_{\bf i}\nonumber \\ & & - V\sum_{{\bf i},\delta}(c^\dagger_{{\bf i}+\delta\up}c^\dagger_{{\bf i}\down}+c^\dagger_{{\bf i}\up}c^\dagger_{{\bf i}+\delta\down})\Delta^\delta_{\bf i}+{\mbox {H.c.}}
\label{eq:BCS}
\end{eqnarray}
where $t$ - the energy unit - is the hopping amplitude for electrons $c_{{\bf i}\sigma}$ on n.n. sites. 
$\mu$, the chemical potential, controls the particle density $\langle$$n$$\rangle$=1/$N$$\sum_{\bf i}$$n_{\bf i}$,
$\delta$=$\pm$${\bf x,y}$ denotes n.n. on an $N$=$L$$\times$$L$ lattice, and in the standard derivation
$\Delta^\delta_{\bf i}$=$\langle c_{{\bf i}\downarrow}
c_{{\bf i}+\delta\uparrow}\rangle$ ($\lag$...$\rag$ signals thermal averaging). In the
usual MF approach to Eq.(\ref{eq:BCS}) the gap function $\Delta^\delta_{\bf i}$ is assumed
a real number, but here we retain the degrees of freedom associated with the
phases and therefore write
$\Delta^\delta_{\bf i}$=$|\Delta_{\bf i}|\exp(i\phi^\delta_{\bf i})$.
The amplitudes are regarded as site variables, whereas the phases are treated as link variables.
$V$($>$0), the n.n. attraction, is assumed to be constant throughout the lattice, but
inhomogeneous generalizations can be implemented in a straightforward manner.  
To calculate observables one needs to determine the corresponding partition function $Z_{\rm SC}$
at temperature $T$=1/$\beta$,
\be
Z_{\rm SC} = \prod_{{\bf i}=1}^{2N} \int_0^{\infty} \hspace{-0.2cm}d|\Delta_{\bf i}| Z_{\rm cl}\int_0^{2\pi} \hspace{-0.2cm} d\phi^x_{\bf i} d\phi^y_{\bf i} Z_c(\{|\Delta_{\bf i}|\}, \{ \phi^{x,y}_{\bf i} \}),
\label{eq:ZCS}
\ee
which is calculated via a canonical MC integration over both
$|\Delta_{\bf i}|$ and the phases $\{\phi^{x,y}_{\bf i}\}$. The electronic partition function
$Z_{\rm c}$=Tr $\{ e^{-\beta H_{\rm SC}'(|\Delta_{\bf i}|, \phi^{x,y}_{\bf i} )} \}$
($H_{\rm SC}'$ being the purely fermionic part of (\ref{eq:BCS})) is obtained after exactly diagonalizing $H_{\rm SC}'$ for
a given fixed set of $|\Delta_{\bf i}|$'s and $\{\phi^{x,y}_{\bf i}\}$ and
finding the eigenvalues $E_n'$; it is then calculated in a standard fashion as
$Z_c$=$\prod_{n=1}^{2N}(1+\exp(-\beta E_n'))$ \cite{foot_ek}. The classical part of $Z_{\rm SC}$ is $Z_{\rm cl}$=$e^{(-4\beta V)(\sum_{{\bf i}}|\Delta_{\bf i}|^2)}$.
The most CPU-time consuming task is the diagonalization leading to the eigenvalues $E_n'$ for a given
set of classical fields, limiting the lattice size.
The results presented here were obtained for lattices up to $N$=14$\times$14, and for temperatures
as low as $T$=0.002$t$. Observables such as the spectral function $A({\bf k},\omega)$, $N(\omega)$=$\sum_{\bf k}$$A({\bf k},\omega)$ 
or the optical conductivity $\sigma$$(\omega$) can be calculated straightforwardly \cite{Dagotto_1}. 
Here, however, we focus on other quantities of particular interest, namely the phase
correlation function $S$(${\bf l}$=($l_x$,$l_y$))=$\frac{1}{N}$$\sum_{\bf i}$$\langle e^{i\phi^x_{\bf i}}e^{-i\phi^x_{{\bf i}+{\bf l}}}\rangle$,
the ``mixed'' correlation $F$(${\bf l}$)=$\frac{1}{N}$$\sum_{\bf i}$$\langle e^{i\phi^x_{\bf i}}e^{-i\phi^y_{{\bf i}+{\bf l}}}\rangle$, which
is determined by the internal symmetry of the pairing electrons as shown below \cite{foot_2}, and the gap
$\Delta_{\rm MC}$$\equiv$$\frac{1}{N}$$\sum_{\bf i}$$\langle$$|$$\Delta_{\bf i}$$|$$\rangle$.
 
\begin{figure}
\centerline{\includegraphics[width=8.0cm,angle=0]{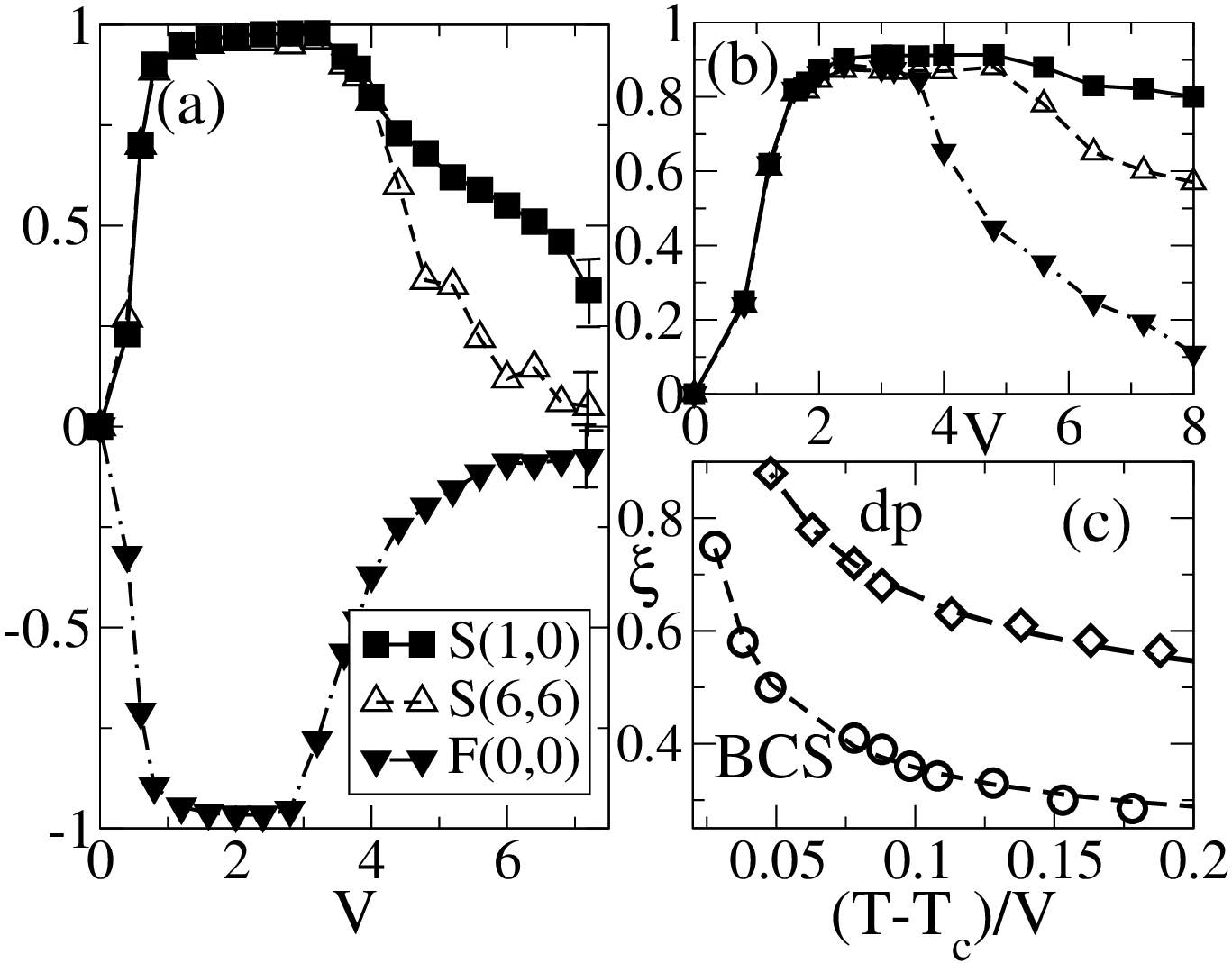}}
\caption{ (a) The phase correlation functions $S$($V$), $F$($V$) at the shortest and maximum (linear) lattice distance at low $T$ 
and $\lag n \rag$=1.
(b) Same functions as in (a), now for $\lag n \rag$$\sim$0.18, leading to (extended) $s$-wave behavior.
The symbols stand for $S$(1,0) (filled squares), $S$(5,0) (open triangles) and $F$(0,0) (filled triangles).
The statistical error is much smaller than the symbols for $V$$<$4, and roughly the symbol
size for $V$$\lesssim$6. (c) The correlation length $\xi$($T$) for $V$=5.6 at $\langle$$n$$\rangle$=1 from the $H_{\rm SC}$ model. 
From the Kosterlitz-Thouless fit (broken line) we obtain $T_c$$\simeq$0.08. Also shown are results for the d-p model at $V$=4.8, that lead to 
$T_c$$\simeq$0.17.} 
\label{Figure_1}
\vspace*{-0.2cm}
\end{figure}

Figure \ref{Figure_1}(a) shows $S$($i_x$,$i_y$), 
$F$(0,0) on a 12$\times$12 lattice, at $\langle$$n$$\rangle$=1 and $T$=0.01.  
The different regimes emerging as $V$ is increased can easily be identified:
(i) a BCS phase extending up to $V$$\simeq$3, where the correlation between
n.n. sites (${\bf i}$, ${\bf j}$=${\bf i}$+${\bf x}$) and (${\bf j}$=${\bf i}$+($L$/2,0))
is virtually identical,
(ii) an intermediate region 3.5$\lesssim$$V$$\lesssim$6,
and (iii) 
the strongly coupled regime $V$$\gtrsim$6, with $short$-$range$ (SR) 
$phase$ $correlations$ only, at least at the lowest temperatures of our simulations.
For the BCS state, $F$(0)$\approx$-1, equivalent to 
$\langle$$\Delta^x_{\bf i}$$\rangle$$\simeq$-$\langle$$\Delta^y_{\bf i}$$\rangle$,
clearly exposing the $d_{x^2-y^2}$-character caused by strong scattering for the Fermi surface (FS) points ($\pm$$\pi$,0), (0,$\pm$$\pi$). This regime is characterized by a unique {\it global} phase, with only thermal fluctuations  
(and finite size effects) responsible for the small deviations from a perfect dSC \cite{foot_a}.
This is revealed by the phase histograms,  which feature two well-defined Gaussian curves centered around $\phi^x_0$ and 
$\phi^y_0$=$\phi^x_0$+$\pi$, respectively.
On the other hand, the until now unexplored strong-coupling regime of Eq.(1) is characterized
by $distributions$ with multiple peaks and no evident global phase \cite{foot_ps}. Such complicated distributions appear 
irrespective of starting configurations and other details of the MC process.  
Below, we will work with $H_{\rm SC}$ as well as with a $d$-wave-projected (``d-p'') 
model where $\Delta^y_{\bf i}$$\equiv$-$\Delta^x_{\bf i}$ is enforced, 
a commonly used {\it approximation} for dSC. 

To further explore the validity of the MC integration we have also performed calculations in the low-density
limit, $\langle$$n$$\rangle$$\sim$0.18, with results for $S$(${\bf l}$), $F$(${\bf l}$) presented in Fig.\ref{Figure_1}(b). 
Here, $F$(0,0)$\approx$1 emerges naturally ($T$ small),
and therefore the expected $d_{x^2+y^2}$-symmetry ($\equiv$$s^\star$, $\cos{k_x}$+$\cos{k_y}$) is realized. 
Again, it is possible to differentiate between weak- and strong-coupling regimes,
based on the same arguments as in (a). 

$\Delta_{\rm MC}$($T$=0.01,$\langle$$n$$\rangle$=1) is shown in the table below for both 
$H_{\rm SC}$ and the d-p model ($\equiv$$\Delta_{\rm MC, dp}$). For small $V$, $\Delta_{\rm MC}$
barely deviates from its MF (dSC) value $\Delta_{\rm MF}$, but it is decidedly 
{\it larger than} $\Delta_{\rm MF}$ in the strongly fluctuating regime \cite{comment_gap}.
This, together with the results shown in Fig.\ref{Figure_1}(a), where $F$(0,0) is very different from -1, signals 
the gradual transition from a dSC into what should be a $s^\star$+$i$$d$-SC \cite{Micnas_1}. 
The SC properties for large $V$ are not so much dictated by the 
FS topology (and band-filling) any more; instead the interaction $V$ forces all electronic states to take part in the 
pairing and not just the ``preferred'' ones near the FS, driving the system away from the $d$-wave state. 
Such a transition, believed to appear for any non-sSC, 
can easily be overlooked in studies biased towards dSC.
\begin{table}[hbtp]
\caption{Comparison of $\Delta_{\rm MC}$ ($\Delta_{\rm MC,dp}$) with the MF gap function, for both $d$- and conventional $s$-wave. \vspace{0.1cm}}        
\centering
\begin{tabular}{lc|cccccccccc}
    V  & &  $\Delta^{10\times10}_{\rm MF}$  & & & $\Delta^{10\times10}_{\rm MF,s}$ & & & $\Delta^{10\times10}_{\rm MC,dp}$ & & & $\Delta^{10\times10}_{\rm MC}$ \\ \hline
   1.2 & & 0.322   & & & 0.241 & & & 0.32$\pm$0.02 & & & 0.32$\pm$0.02  \\        
   2.0 & & 0.627   & & & 0.666 & & & 0.62$\pm$0.02 & & & 0.62$\pm$0.03  \\        
   4.0 & & 1.420   & & & 1.747 & & & 1.42$\pm$0.03 & & & 1.63$\pm$0.18  \\        
   4.8 & & 1.743   & & & 2.213 & & & 1.90$\pm$0.20 & & & 2.05$\pm$0.18            \\
   5.6 & & 2.066   & & & 2.636 & & & 2.50$\pm$0.20 & & & 2.50$\pm$0.20    \\ \hline\hline 
\end{tabular}
\label{table_1}
\vspace*{-0.1cm}
\end{table}
For $V$$\gg$1, $\Delta_{\rm MC}$ is slightly smaller than the MF gap for the simple s-wave state, 
$\Delta_{\rm MF,s}$; thus, in this regime both the $d$-and the $s^\star$-wave gap will have almost the same amplitude \cite{Kotliar_1} and it 
may resemble a disordered sSC. Because of its $s$-wave component, the resulting state has a nodeless 
FS, i.e. no gapless excitations, and thus is strinkingly different from the weak-coupling state.  
 
\begin{figure}
\vspace{-0.2cm}
\centerline{\includegraphics[width=8.2cm,angle=0]{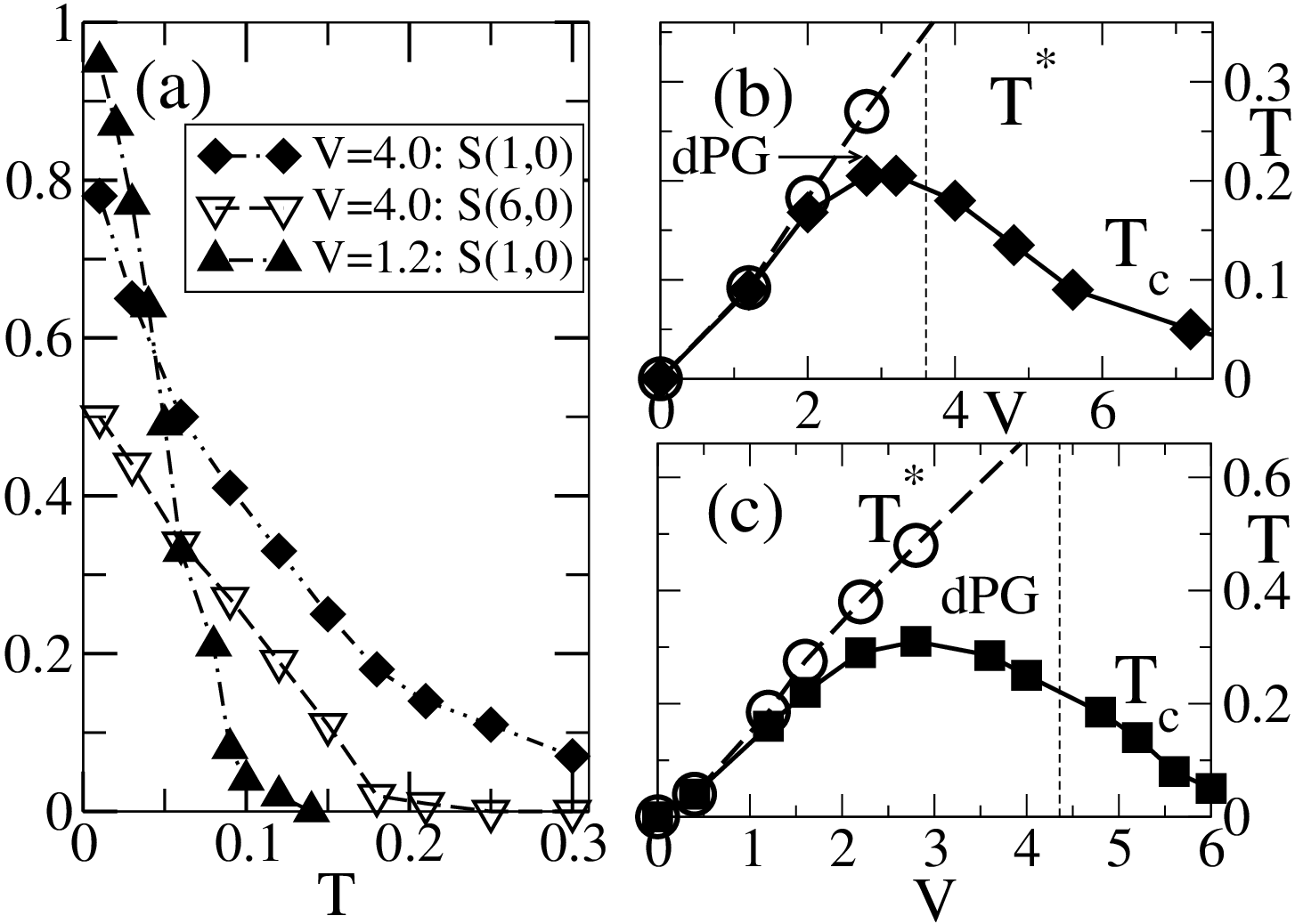}}
\caption{(a) SR and LR phase correlation functions vs. $T$ for two different values of $V$,
covering the weak- and strong-coupling regime, for $\lag n \rag$=1.
Once $V$$\lesssim$2, no difference between short- and long-range correlations is observed, and $S$(6,0) ($V$=1.2) is not shown for reasons of clarity. 
Symbol sizes roughly match the errors. (b) The phase diagram for $H_{\rm SC}$ derived from (a); $T_c$ and $T^\star$ as explained in the text. 
(c) shows the phase diagram for the d-p model. Note the differences 
between (b) and (c). The $d$-wave PG regime (dPG) is indicated in both (b), (c). Beyond the dashed lines the $d$-wave character of $H_{\rm SC}$ is lost, whereas the d-p model incorporates higher harmonics as well.    
}
\label{Figure_2}
\vspace*{-0.3cm}
\end{figure}

The investigation of the temperature dependence of $S$(${\bf l}$) for both $H_{\rm SC}$ and the d-p model allows us to introduce for the first time in a BCS-like Hamiltonian two 
characteristic temperatures $T^\star$ and $T_c$ in the case of strong coupling, in contrast to the BCS regime, where this distinction does not exist. 
We associate $T^\star$ with the temperature where SR phase correlations develop
(defined here as $S(1,0)$$\geq$0.1, but other cutoffs lead to quite similar qualitative
conclusions). On the other hand, 
 $T_c$ is commonly identified with the onset of long-range 
phase coherence (here we use the criterium $S(L/2,0)$$\geq$0.1). 
$T^\star$ and $T_c$ are essentially identical for $V$ not too large (Fig.\ref{Figure_2}(a)), 
and they are only clearly different for $V$$\gtrsim$3, with $T^\star$ larger than $T_c$ by a factor of 3-4 for $V$$>$5 
(Fig.\ref{Figure_2}(a)) \cite{foot_lowden}.  
Based on such MC results, a phase diagram, presenting $T_c$ and $T^\star$ as a function of the pairing
attraction, is displayed in Figs.\ref{Figure_2}(b), (c). Remarkably, the values of $T_c$ reach a maximum 
$T^{max}_c$$\simeq$ 0.2 for $V_{\rm max}$$\approx$3 
(similar to other such reported values), whereas $T^\star$ increases steadily with $V$ \cite{commentUV}. 
For the ``rigid'' projected model (Fig.\ref{Figure_2}(c)), $T^{\max}_c$$\simeq$0.3 
, accompanied by a more prominent regime of SR correlations. The regime of a $d$-wave PG (dPG) is indicated, 
and although it is sizeable for the d-p model, it is a rather small window for the more realistic $H_{\rm SC}$. For the latter model, {\it the state with 
a large difference between $T^\star$ and $T_c$ (typical for HTS) is only found for values of $V$ 
that do not lead to a dSC at low $T$}, nowadays widely accepted for HTS, owing to strong experimental evidence. 
For $\langle$$n$$\rangle$$<$1, the dPG should be even less prominent than shown in Fig.\ref{Figure_2}(b). 
This disagreement between theory and experiment puts the thermal phase-fluctuation scenario for HTS into serious doubt. 
$T_c$ itself is a continous function (Fig.\ref{Figure_2}(b),(c)), smoothly connecting the limits 
$V$$\rightarrow$0,$\infty$, as predicted in an early work \cite{Schmitt_Rink_1}. Although the existence of
$T^{\rm max}_c$ has long been known, this is - to our knowledge - 
the first time it has been directly established in the framework of $H_{\rm SC}$, 
since self-consistent methods are tracking $T^\star$ rather than $T_c$ \cite{foot_Ts}. Yet, as demonstrated in Fig.\ref{Figure_2},
they work very well for $V$ not excessively large.  

For $V$$\gtrsim$$V_{\rm max}$ one presumably enters the realm of pronounced Kosterlitz-Thouless (KT) physics \cite{Kosterlitz_1},
whence $T_c$ is dictated by vortex binding rather than Cooper pairing. 
The critical temperature $T_{\rm KT}$$\equiv$$T_c$ in such models is proportional to 1/$V$, following a perturbative analysis,
similar to what is found in Fig.\ref{Figure_2}(b),(c). 
It is certainly non-trivial to establish whether or not KT-behavior is found for $H_{\rm SC}$, which, unlike the 
standard $X$$Y$ model, couples fermions to classical fields.  
For this purpose, we extract a correlation length $\xi$ by fitting $S$(${\bf l}$) with an exponential, 
$S$($r_x$)$\propto$$\exp(-r_x/\xi)$, and explore its temperature dependence, which should behave as
$\xi$($T$)$\propto$$\exp$[A/$\sqrt{T-T_c}$]. In the case of $V$=5.6, such a KT analysis (for 0.10$\leq$$T$$\leq$0.35) produces a very good fit for
$\xi$($T$) (see Fig.1(c)) and yields $T_c$=0.08$\pm$0.01,
remarkably close to what has been established with our alternative 
definition of $T_c$ above. In addition, the exponential fit
is not possible for $T$$\lesssim$0.08 - signalling that $H_{\rm SC}$ is entering a state with different scaling behavior.
In a similar fashion, $T_c$ is found to be 
0.17$(\pm$0.02) for the d-p model at $V$=4.8, only slightly lower than its estimate from $S$(${\bf l}$). 
Although the precise values cited above - having been obtained on relatively small lattices - need to be 
cautiously considered, our results are compatible with 
KT physics governing the region between $T_c$ and $T^\star$, even in the presence of fermions.

\begin{figure}
\centerline{\includegraphics[width=7.5cm,angle=0]{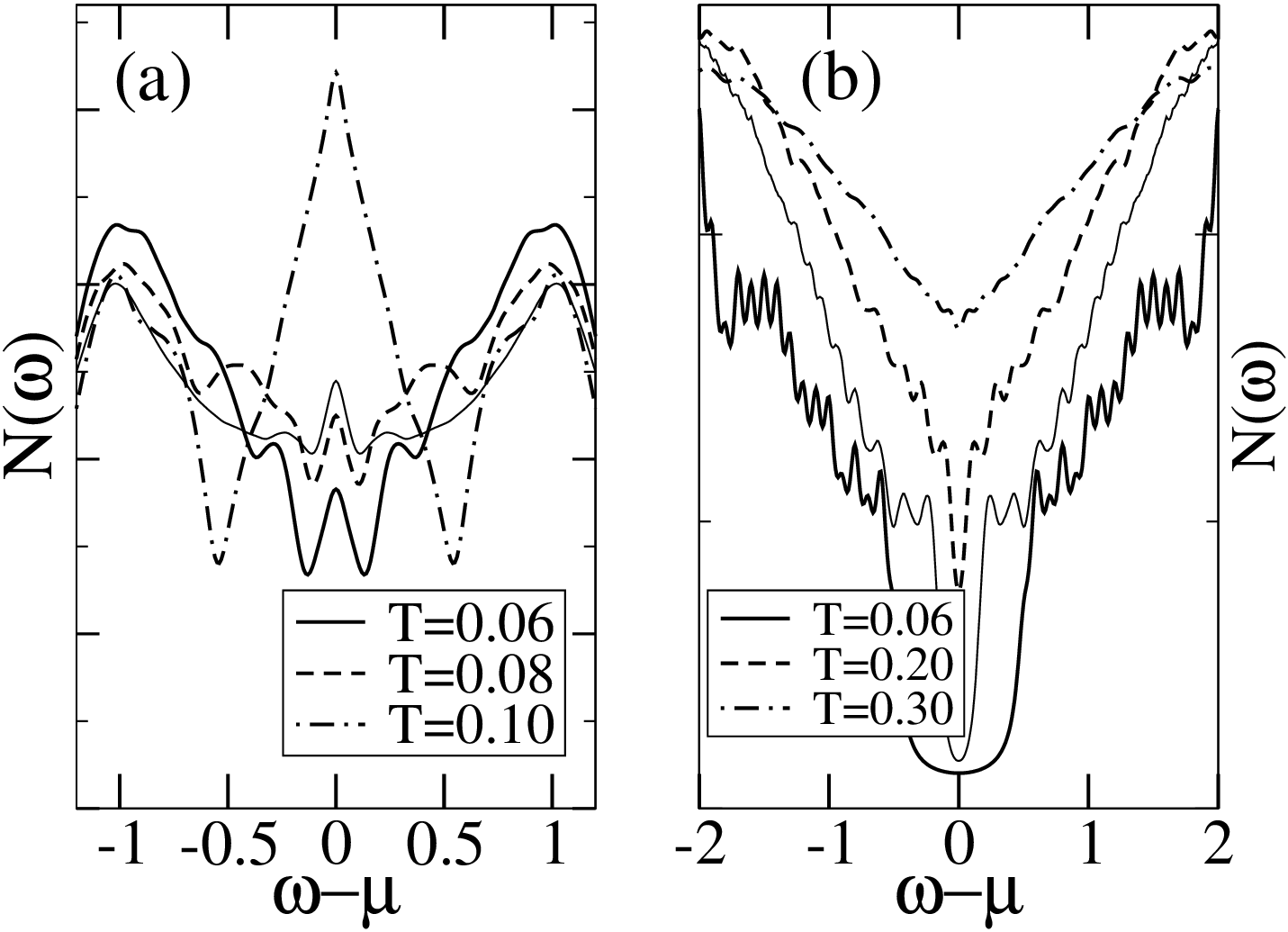}}
\caption{The $T$-dependent $N$($\omega$) (12$\times$12) for weak- ((a), $V$=1.2) and intermediate-coupling ((b), $V$=4) for $H_{\rm SC}$,
illustrating the development of the PG as $T$ is lowered. In (a) the system is uncorrelated just above $T_c$, whereas in (b) the PG appears at $T^\star$
concurrently with SR ordering. For $T$=0.06, the deviation from the $d$-wave state becomes evident, and a gap $\omega_0$$\simeq$0.6 opens up.
A broadening $\gamma$=0.05 was used. For reasons of comparison, the two thin lines represent the d-p model for T=0.15 ($V$=1.2), and T=0.20 ($V$=4), 
respectively.}
\label{Figure_3}
\vspace*{-0.2cm}
\end{figure}

As the PG scenario would suggest, the effect of $T^\star$ is clearly visible in $N$($\omega$), 
which is shown for $V$=1.20 and $V$=4.0 in Fig.\ref{Figure_3}.
In the BCS limit (a), a gap appears for temperatures $T^\star$$\simeq$$T_c$=0.09 (compare to Fig.\ref{Figure_2}(b)),
whereas $N$($\omega$) shows non-correlated behavior and a van-Hove peak just above $T_c$. The remaining small peak 
at $\omega$=0 ($T$$<$$T_c$) is a finite size effect. At $V$=4.0 (Fig.\ref{Figure_3}(b)), however, there is a wide region below $T^\star$$\simeq$0.30 
(coinciding with the onset of SR fluctuations (Fig.\ref{Figure_2}(b))
where a PG exists in $N$$(\omega)$ without LRO in the phase correlations.
As $T$ is lowered, spectral weight is continously removed from small energies, and the dip centered around $\omega$$\sim$0 deepens.
$N$($\omega$) has a {\it true} gap at lower
$T$ (finite spectral weight at $\omega$$\approx$0 stems from broadening only), reflecting the deviations from the $d_{x^2-y^2}$-symmetry noted 
before. In contrast, the projected model has a $d$-wave-like gap above $T_c$ (Fig.\ref{Figure_3}(b)).    
Finite size effects in general influence subtle signals such as $d$-wave gaps considerably, but the observations above strongly validate our definition of
$T^\star$ and demonstrate the influence of SRO on $N$($\omega$), which we have observed for all values of $V$. 

We have also performed 
calculations for a model with diagonal hopping $t'$=-1. For densities $\langle$$n$$\rangle$$\sim$0.2, this produces
electron pockets around ($\pi$,0) (and related points) and, therefore, low-density dSC, confirmed in the same way as shown in
Fig.\ref{Figure_1}(a),(b). Our results can be summarized by 
stating that (a) the BCS region extends to very large $V$$\sim$12, (b) for $V$=10,
it remains in the BCS state even as $\langle$$n$$\rangle$$\rightarrow$0, and (c) $T_c$ decreases concurrently, 
but so does $\Delta_{\rm MC}$,
and, thus would $T^\star$, in disagreement with the well-established phase diagram. In addition, for intermediate $V$ and larger, 
$\mu$ is found to be below the band minimum. 
The resulting absence of nodes in $A$(${\bf k}$,$\omega$) is related to bound-pair formation, as previously noted \cite{Randeria_1}. 
Overall, it seems 
very difficult to reconcile our results here with the observed behavior of the cuprates.

Summarizing, a MC technique has been introduced 
for an unbiased investigation of the SC state as described in the ($d$-wave) pairing Hamiltonian. It
reproduces both the BCS limit as well as the strong-coupling regime 
at all temperatures and densities. 
The establishment of a PG regime in the case of a strong pairing 
between two characteristic temperatures $T^\star$ and $T_c$ 
and an associated non-trivial phase diagram, has been numerically demonstrated.
Our results for $H_{\rm SC}$ seem to indicate that the observed 
PG features of HTS cannot be reconciled with a (classical) phase-fluctuation-dominated dSC.  
The integration method presented here can 
easily be extended to study disordered systems 
as well as to simultaneously investigate the competition of several fluctuation channels, 
such as dSC, antiferromagnetism and charge order \cite{Alvarez_1}, 
in an unbiased fashion. As such, this method
(maybe best dubbed ``mean-field Monte Carlo'') should be an invaluable tool 
in unlocking the secrets of the cuprates and possibly
other systems with strong-coupling aspects such as Bose-Einstein condensates in cold fermions.\nl
Discussions with R. Zeyher, G. Khaliullin, P. Horsch, M. Randeria, R. Micnas and W. Metzner are gratefully acknowledged. 
E.D. is supported by grant NSF-DMR 0443144. \vspace*{-0.4cm}

\bibliographystyle{unsrt}



\begin{thebibliography}{20}
\bibitem{foot_2} If only the phases are of interest, $|\Delta|$ may be assumed constant
(using, e.g., $\Delta_{\rm MF}$ instead) and the integration performed over $\{ \phi^{x,y}_{\bf i} \}$ only.
\bibitem{comment_gap} The curious transition region is characterized by slow MC thermalization and convergence for $\Delta_{\rm SC}$.
\bibitem{Micnas_1} R. Micnas {\it et al.}, J. Phys. Cond. Matter {\bf 14}, 9631 (2002). 
This is also confirmed by an analysis of the respective energies as well as by snapshots that show $\Delta^x_{\bf i}$$\simeq$$i$$\Delta^y_{\bf i}$.
\bibitem{foot_lowden} This works in the low-$\langle$$n$$\rangle$-limit, too. We find, e.g. $T_c$=0.08 ($V$=4, $\langle$$n$$\rangle$$\sim$0.2), 
so $T_c$ is smaller in this limit, as expected.
\bibitem{commentUV} A similar phase diagram (with smaller $T_c$'s) 
can be found also in the t-U-V model ($\langle$$n$$\rangle$=1) for intermediate values of 
U, where both AF and SC coexist.
\end{thebibliography}

\newpage





\end{document}